\documentclass[preprint,prd,aps,amsmath,superscriptaddress,nofootinbib,tightenlines]{revtex4}

\usepackage[hypertex]{hyperref}
\usepackage{graphicx}
\usepackage{epstopdf}
\usepackage{bm}
\usepackage{epsfig}
\usepackage{xspace}

\def\OMIT#1{}

\newcommand{\nn}{\nonumber} 

\newcommand{\bn}{{\bar n}}
\newcommand{\bea}{\begin{eqnarray}}
\newcommand{\eea}{\end{eqnarray}}

\newcommand{\mcdot}{\!\cdot\!}

\newcommand{\plus}{\ensuremath{\! + \!}}
\newcommand{\minus}{\ensuremath{\! - \!}}

\def\lsim{\mathrel{\raise.3ex\hbox{$<$\kern-.75em\lower1ex\hbox{$\sim$}}}}
\def\gsim{\mathrel{\raise.3ex\hbox{$>$\kern-.75em\lower1ex\hbox{$\sim$}}}}

\begin{document}
\setlength\baselineskip{17pt}


\preprint{
 \vbox{ \hbox{MIT-CTP 3792} 
 \hbox{MPP-2007-135}  
 \hbox{arXiv:0709.3519} }}

\title{Designing Gapped Soft Functions for Jet Production} 

\vspace*{1cm}

\author{Andre H.~Hoang}
  \affiliation{Max-Planck-Institut f\"ur Physik (Werner-Heisenberg-Institut) 
  F\"ohringer Ring 6, M\"unchen, Germany, 80805 \footnote{Electronic address: 
 ahoang@mppmu.mpg.de}}

\author{Iain W.~Stewart\vspace{0.4cm}}
  \affiliation{Department of Physics, Massachusetts Institute of Technology, 
Boston, MA 02139 \footnote{Electronic address: iains@mit.edu}\vspace*{0.5cm}}



\begin{abstract}
\vspace{0.4cm}

Distributions in jet production often depend on a soft function, $S$,
which describes hadronic radiation between the jets. Near kinematic thresholds
$S$ encodes nonperturbative information, while far from thresholds $S$ can be
computed with an operator product expansion (OPE). We design soft functions for
jets that serve this dual purpose, reducing to the perturbative result in the
OPE region and to a consistent model in the nonperturbative region.  We use the
$\overline{\rm MS}$ scheme, and in both regions $S$ displays the appropriate
renormalization group scale dependence.  We point out that viable soft function
models should have a gap associated with the minimum hadronic energy deposit.
This gap is connected to the leading ${\cal O}(\Lambda_{\rm QCD})$ renormalon
ambiguity in jet event shapes.  By defining the gap in a suitable scheme we
demonstrate that the leading renormalon can be eliminated.  This improves the
convergence of perturbative results, and also the stability by which
non-perturbative parameters encode the underlying soft physics.

\end{abstract}

\maketitle

\newpage

\newpage

Soft functions play an important role in the study of cross sections close to
kinematic thresholds, characterized by jets of collimated hadrons with small
invariant mass. These cross sections are frequently described by factorization
theorems involving hard Wilson coefficients, jet functions describing the jets
of hadrons, and a soft function $S$. The hard coefficients and the jet functions
are perturbative, while $S$ encodes universal nonperturbative information on
soft radiation between the jets. The prototype examples are event-shape
distributions in $e^+e^-$ annihilation for large c.m.\,\,energies
$Q$~\cite{Korchemsky:1998ev,Korchemsky:1999kt,Bauer:2003di}, such as the thrust
$T$~\cite{Catani:1991kz,Korchemsky:1994is,Dokshitzer:1997ew}, where $T\equiv{\rm
  max}_{\hat n} \sum_i |\vec p_i\cdot \hat n| / \sum_i |\vec
p_i|$~\cite{Farhi:1977sg} and the kinematically allowed range is $1/2<T<1$. In
the threshold ``dijet'' region of large thrust, $T\sim 1$, the events are
characterized by two back-to-back jets, and at leading order in $1/Q$ the
factorization theorem has two jet functions and one soft
function~\cite{Korchemsky:1998ev,Korchemsky:1999kt,Bauer:2003di}.  Other
examples include distributions for jet broadening~\cite{Catani:1992jc}, the
heavy jet mass~\cite{Chandramohan:1980ry}, and their generalization to
angularities~\cite{Berger:2002ig}. The dijet region also plays a crucial role in
event shapes for massive particles, such as the invariant mass distribution of
jets from top-quarks~\cite{Fleming:2007qr}. For applications at hadron colliders
soft functions which account for initial state radiation are
important~\cite{Kidonakis:1998ur}.  Finally, for studies of weak $B$-meson
decays to jets, soft functions involving the initial state $B$ play a crucial
role.  Examples are $B\to X_s\gamma$ and $B\to X_u
e\bar\nu$~\cite{Neubert:1994um,Bigi:1994ex,Mannel:1994pm}, as well as $B\to
X_s\ell^+\ell^-$~\cite{Lee:2005pk,Lee:2005pw}. Here phase space cuts enhance the
region where the soft function has a large effect.

Near threshold one can distinguish two regions.  Very close to threshold the
distribution typically shows an enhanced peaked structure, and nonperturbative
information in the soft function is important for determining the shape and the
maximum of the distribution. The size of this ``peak region'' is set by the
hadronic scale $\Lambda_{\rm QCD}$. Next to the peak region the distribution
typically falls off and shows a tail-behavior but is not yet highly suppressed.
The dynamics is still dominated by jets and soft radiation, but in this ``tail
region'' the leading soft function can be computed perturbatively since it is
probed at scales larger than $\Lambda_{\rm QCD}$. In the tail region operators
sensitive to nonperturbative physics are power-suppressed.  Computations of
moments involving integrations over both the peak and tail regions can be done
with this same power expansion.

Since $S$ encodes different types of physics in the peak and in the tail region,
one possibility is to make separate predictions for the corresponding
cross-sections.  However, phenomenologically it is often desired to treat both
regions coherently. In a pioneering analysis of $e^+e^-\to {\rm
  jets}$~\cite{Korchemsky:2000kp} this was handled by implementing a ``hard'' IR
cutoff on the event shape variable ``$e$''. Perturbation theory was used above
the cutoff and the perturbative corrections were frozen below it, with
\begin{align}
 R_{\rm PT}(e,\Lambda_{\rm IR}) = \theta\Big(e-\frac{\Lambda_{\rm IR}}{Q}\Big) R_{\rm
   PT}^{\rm NLL}(e) + \theta\Big(\frac{\Lambda_{\rm IR}}{Q}-e\Big) R_{\rm
   PT}^{\rm NLL}(\Lambda_{\rm IR}/Q) \,,
\end{align}
where $R_{\rm PT}^{\rm NLL}$ contained perturbative results up to two-loop order
with next-to-next-to-leading log resummation (NLL).  The function $R_{\rm
  PT}(e,\Lambda_{\rm IR})$ was then convoluted with a normalized soft function
model $S_{\rm mod}$ as dictated by the factorization theorem. With a simple
choice for $S_{\rm mod}$ good agreement with LEP data was found for several
event shapes.  This cutoff procedure does not attempt to treat explicitly the
renormalization scale dependence in the region where the soft function is
non-perturbative, nor does it systematically implement the
perturbative corrections in this peak region.

The multi-region issue has also been analyzed in the context of $B$-meson
decays.  In Ref.~\cite{Bosch:2004th} a perturbative tail was glued to the
soft-function model,
\begin{align}
\label{Sglued}
  S(\hat\omega,\mu) = S_{\rm mod}(\hat\omega) +
  \theta(\hat\omega-\Lambda-\mu/\sqrt{e}) S_{\rm part}(\hat\omega,\mu) \,,
\end{align}
where $S_{\rm part}$ is the ``partonic'' soft function obtained from
perturbation theory and where the argument in the $\theta$-function was chosen
such that the tail turns on without discontinuity, using the condition $S_{\rm
  PT}(\hat\omega,\sqrt{e}(\hat\omega-\Lambda))=0$.  This method provides the
correct renormalization group behavior for the treatment of the tail region at
leading order, and is an improvement because it allows the perturbative jet
function corrections to be incorporated systematically in the peak region.
Shortfalls are that in the peak region it still hides the dependence on the
renormalization scale $\mu$ in model parameters, and that the perturbative tail
is turned on by hand at a particular point, rather than allowing it to appear
once it dominates the non-perturbative corrections.

In this paper we develop a procedure for constructing soft function models for
jets that i) reduce to the perturbative result in the OPE region and a
consistent model in the nonperturbative region, ii) exhibit the proper
renormalization group scale dependence in the $\overline {\rm MS}$ scheme, iii)
have a gap associated with the minimum hadronic energy deposit, and iv) are
stabilized to perturbative corrections by being free from the leading ${\cal
  O}(\Lambda_{\rm QCD})$ renormalon. We show that the soft function gap
parameter is essential for removing the renormalon ambiguity of the partonic
threshold energy order-by-order in perturbation theory.

Although our procedure is quite general, in order to make all the steps explicit
we will carry it out in the context of a specific example. We consider event
shapes for top-quark jets produced in $e^+e^-\to t\bar t$ at c.m.~energies $Q
\gg m_t$. The soft function we construct applies equally well for massless event
shapes in the dijet region, that is, very little of our discussion depends on
the presence of the top-quark mass or width.  We consider the double
differential top-antitop invariant mass distribution, $d^2\sigma/dM_t dM_{\bar
  t}$, where $M_{t,\bar t}$ are either in the peak or the tail region.  In the
peak region near the top mass resonance, $s_{t,\bar t}\equiv (M_{t,\bar
  t}^2-m_t^2) \sim m_t \Gamma_t$ where $\Gamma_t$ is the top-quark width, and we
have the factorization theorem~\cite{Fleming:2007qr}
\begin{align} \label{FactThm}
  \frac{d\sigma^{\rm peak}}{ dM_t^2\, dM_{\bar t}^2} &= 
  \sigma_0 \: H(Q,m_t,\mu)\!
  \int\! d\ell^+ d\ell^- B_+\Big(\frac{s_t - Q\ell^+}{m_t},\mu\Big)\:
  B_-\Big(\frac{s_{\bar t} - Q\ell^-}{m_t},\mu\Big) 
  S_{\rm np}(\ell^+\!,\ell^-\! ,\mu) \,,
\end{align}
which is valid at leading order in $m_t/Q$ and $\Gamma_t/m_t$.  Here $H$ is a
calculable hard coefficient and $B_\pm$ are calculable jet functions, whereas
$S_{\rm np}(\ell^\pm,\mu)$ is a nonperturbative soft function which peaks for
$\ell^\pm\sim \Lambda_{\rm QCD}$ when $\mu\sim\ell^\pm$.  In general the
convolution probes momenta $\ell^\pm \sim s_{t,\bar t}/Q$ in the soft function,
and large logs in $S$ are avoided by taking $\mu\sim \ell^\pm$ and summing large
logs in the jet and hard functions.  In the peak region $s_{t,\bar t}\sim
Q\Lambda_{\rm QCD} + m_t\Gamma_t$, so the nonperturbative distribution described
by $S_{\rm np}(\ell^\pm,\mu)$ directly effects the differential cross section.
On the other hand, in the tail region, $ s_{t,\bar t} \gg Q\Lambda_{\rm QCD} +
m_t\Gamma_t$, and the dominant momenta in the soft function are $\ell^\pm \sim
s_{t,\bar t}/Q$. In the interesting region this is a perturbative scale of
$\ell^\pm\simeq 3-30\,{\rm GeV}$ or larger, depending on the size of $Q$. The
leading order factorization theorem in this tail region is
\begin{align} \label{SFactThm}
 \frac{d\sigma^{\rm tail}}{ dM_t^2\, dM_{\bar t}^2} &= 
  \sigma_0 \: H(Q,m_t,\mu)\!
  \int\! d\ell^+ d\ell^- B_+\Big(\frac{s_t \minus Q\ell^+}{m_t},\mu\Big)\:
  B_-\Big(\frac{s_{\bar t} \minus Q\ell^-}{m_t},\mu\Big) 
  S_{\rm part}(\ell^+,\ell^-,\mu) \,,
\end{align}
which is valid to leading order in $s_{t,\bar t}/Q^2$, $m_t/Q$, and
$\Lambda_{\rm QCD} Q/s_{t,\bar t}$. Here the partonic soft function $S_{\rm
  part}(\ell^\pm,\mu)$ can be computed as a perturbative series in $\alpha_s$.
Power corrections at ${\cal O}(\Lambda_{\rm QCD} Q/s_{t,\bar t})$ are determined
from $S_{\rm np}$ in a manner discussed below, while power corrections at ${\cal
  O}(s_{t,\bar t}/Q^2)$ involve new factorization theorems containing subleading
soft functions (which have been worked out for inclusive
$B$-decays~\cite{Bauer:2002yu,Lee:2004ja,Bosch:2004cb,Beneke:2004in}).

The soft function carries information on how soft radiation is associated to the
definition of the invariant mass variables $M_{t,\bar t}$.  To be definite we
consider hemisphere mass definitions where the soft function for both
Eqs.~(\ref{FactThm}) and (\ref{SFactThm}) is~\cite{Fleming:2007qr}
\begin{align} \label{Shemi}
  S(\ell^\pm,\mu) &\equiv \frac{1}{N_c}\sum _{X_s} \delta(\ell^+\! \minus
  k_s^{+a}) \delta(\ell^- \!\minus k_s^{-b}) \langle 0| (\overline
  {Y}_\bn)^{cd}\, ({Y}_n)^{ce} (0) |X_s \rangle \langle X_s|
  ({Y}^\dagger_n)^{ef}\, (\overline {Y}_\bn^\dagger)^{df} (0) |0\rangle .
\end{align}
Here $k_s^{+a}$ is the total plus-momentum of soft hadrons in $X_s$ that are
in hemisphere-a, $k_s^{-b}$ is the total minus momentum for soft hadrons in
the other hemisphere.  The soft function for thrust is related to the
hemisphere soft function by 
\begin{align}
\label{Sthrust}
S_T(\tau)=\int d\ell^+
d\ell^-\delta\Big(\tau-\frac{\ell^++\ell^-}{Q}\Big)\,S(\ell^+,\ell^-)\,,
\end{align}
with $\tau\equiv1-T$, and we emphasize that $S(\ell^+,\ell^-)$ is independent of
the top-mass.  In general soft functions are matrix elements of Wilson lines,
which in our case are
\begin{align} \label{Yn}
  Y_n^\dagger(x) &=   {\rm P} \, 
   \exp\Big(i g\! \int_{0}^\infty \!\!\!\!ds\, n\mcdot A_{s}(ns\!+\!x) \Big) \,,
 & \overline {Y_\bn}^\dagger(x) & =   {\rm P} \: \exp\Big( i g\! \int_{0}^{\infty} \!\!\!\!ds\, 
      \bn\mcdot \overline {A}_{s}(\bn s\!+\! x) \Big) 
  \,.
\end{align}
In order to predict the invariant mass distribution in the peak and the tail
regions we would like to connect Eqs.~(\ref{FactThm}) and (\ref{SFactThm}).  In
this paper we consider the task of constructing an appropriate soft-function
that contains both $S_{\rm np}$ and $S_{\rm part}$ and which can be applied in
the peak and the tail region. In order to be useful the result must remain
consistent for scales $\mu \sim s_{t,\bar t}/Q$, both in the tail region where
$s_{t,\bar t}\gg Q\Lambda$ and in the peak region where $s_{t,\bar t}\sim
m_t\Gamma_t+ Q\Lambda$. We will consider all large logs to have already been
summed by renormalization group evolution from $Q$ down to these $\mu$'s. So the
task is to determine the soft function matrix element at these $\mu$'s, where it
should contain no large logs.

To begin, consider modeling the soft function by
\begin{align}
\label{S1}
S(\ell^+,\ell^-,\mu) & = 
\int_{-\infty}^{+\infty}\!\!\! d\tilde\ell^+
\int_{-\infty}^{+\infty}\!\!\! d\tilde\ell^-\
S_{\rm part}(\ell^+ \minus \tilde\ell^+,\ell^- \minus \tilde\ell^-,\mu)\,
S_{\rm mod}(\tilde\ell^+,\tilde\ell^-) 
\,,
\end{align}
where $S_{\rm part}(\ell^\pm,\mu)$ is the partonic soft function computed in
perturbation theory, and $S_{\rm mod}(\tilde\ell^\pm)$ is a nonperturbative
model function that is $\mu$-independent and contributes only for
$\tilde\ell^\pm \sim \Lambda_{\rm QCD}$.  In Ref.~\cite{Lee:2005pw} an analog to
Eq.~(\ref{S1}) was used in the study of $b\to s\ell^+\ell^-$ to alleviate the
issues mentioned about Eq.~(\ref{Sglued}). Taking $S_{\rm part}$ to ${\cal
  O}(\alpha_s)$ this formula provided a simple way of incorporating the cutoff
OPE moment constraints of Ref.~\cite{Bosch:2004th} in the model for the
nonperturbative $B$-meson soft function.  Here we will argue that, suitably
refined, Eq.~(\ref{S1}) can be used to design soft functions for jets that are
consistent with the desired properties stated earlier.  Defining moments
\begin{align}
  S_{\rm mod}^{[n,m]} \equiv
  \int_{-\infty}^{+\infty}\!\! d\ell^+ d\ell^-\: (\ell^+)^n (\ell^-)^m
  S_{\rm mod}(\ell^+,\ell^-)    \,,
\end{align}
we will demand that $S_{\rm mod}$ is normalized, $S_{\rm mod}^{[0,0]}=1$. We
will also demand that higher moments are finite where we have $S_{\rm
  mod}^{[n,m]}\sim (\Lambda_{\rm QCD})^{n+m}$ for $n+m>0$.

\begin{figure}
\begin{center}
\includegraphics[width=6.4in]{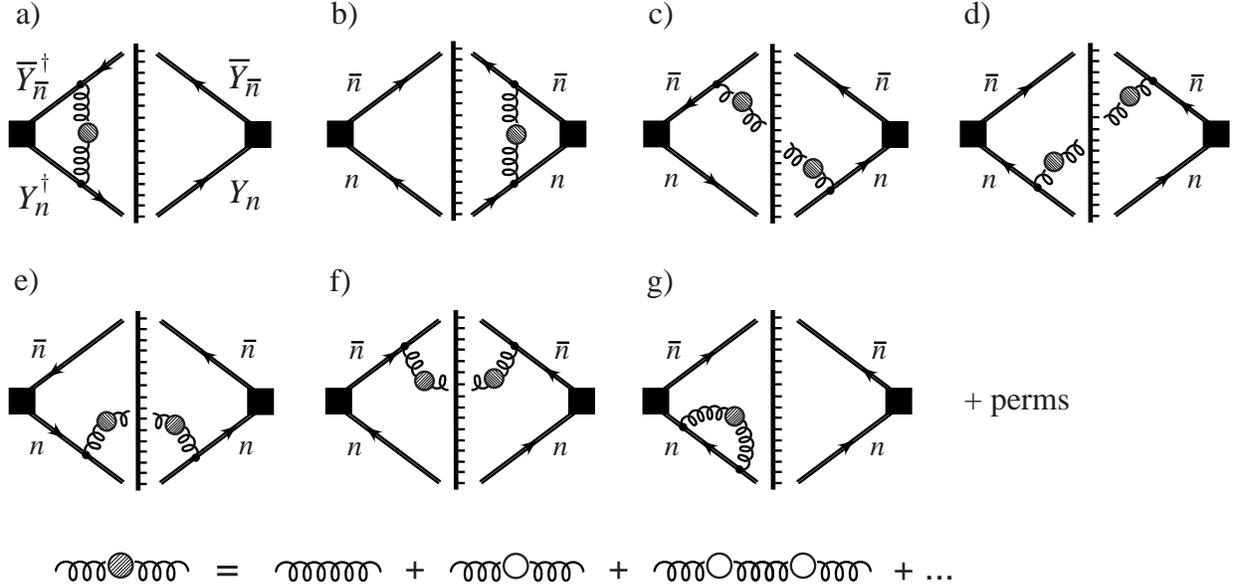}
\vskip-0.2cm
\caption{Graphs for the hemisphere soft function with bubble chains. The
  solid lines denote $Y$-Wilson lines, and the line with ticks is the final
  state cut which may also cut a quark bubble.}
\label{softgraphs}
\end{center}
\end{figure}

A virtue of Eq.~(\ref{S1}) is that it produces by construction the
proper OPE in 
Eq.~(\ref{SFactThm}) when used at a perturbative scale $\mu=\mu_{op}\sim
s_{t,\bar t}/Q\gg \Lambda_{\rm QCD}$ where $\ell^\pm\sim s_{t,\bar t}/Q$. To see
this recall that $\tilde \ell^\pm\sim \Lambda_{\rm QCD}$, and so we can expand
$S_{\rm part}$ for $\tilde \ell^\pm \ll \ell^\pm$ to give
\begin{align} \label{Sop}
 S(\ell^\pm,\mu_{op}) 
 &= S_{\rm part}(\ell^\pm,\mu_{op}) \: S_{\rm mod}^{[0,0]} 
  - \Big[\frac{d}{d\ell^+} S_{\rm part}(\ell^\pm,\mu_{op}) \: S_{\rm mod}^{[1,0]}
  +  \frac{d}{d\ell^-} S_{\rm part}(\ell^\pm,\mu_{op}) \: S_{\rm mod}^{[0,1]}\Big]
 \nn\\
 &\quad  + {\cal O}\Big( \frac{Q^2\Lambda_{\rm QCD}^2}{s^2}\Big) .
\end{align}
Since $S_{\rm mod}^{[0,0]}=1$ we have the desired result that
$S(\ell^\pm,\mu_{op})=S_{\rm part}(\ell^\pm,\mu_{op})$ at leading power.
Computing the renormalized soft function in Eq.~(\ref{Shemi}) to order
$\alpha_s$ (Fig.~\ref{softgraphs} with no $n_f$-bubbles) it factors
as\footnote{We note that the factorized form of the soft function with respect
  to the two hemisphere light-cone variables $\ell^\pm$ in
  Eq.~(\ref{softfactor}) allows for the possibility to choose two different
  $\mu$'s at which to stop running the two jet functions $B_\pm$ in the
  factorization theorems~(\ref{FactThm}) and (\ref{SFactThm}). While we do not
  expect that relation~(\ref{softfactor}) is maintained for non-logarithmic
  corrections beyond the one-loop level, one can prove that the factorized form
  is maintained to all orders as far the scale-dependence is concerned, as in
  Eq.~(\ref{SUS})~\cite{FHMS2}. Thus it is possible to treat the situation where
  $s_t$ and $s_{\bar t}$ are widely separated and to account for the resulting
  non-global logarithms~\cite{Dasgupta:2001sh} by choosing both renormalization
  scales differently. }
\begin{align}
\label{softfactor}
S^{\rm NLO}_{\rm part}(\ell^\pm,\mu)=S^{\rm NLO}_{\rm
  part}(\ell^+,\mu)S^{\rm NLO}_{\rm part}(\ell^-,\mu)
\end{align} 
with
\begin{align} \label{SNLO}
 S_{\rm part}^{\rm NLO}(\ell,\mu)
  = \delta(\ell) + \frac{C_F\alpha_s(\mu)}{\pi} \bigg\{
  \frac{\pi^2}{24}\delta(\ell) - \frac{2}{\mu} \Big[
  \frac{\theta(\ell)\ln(\ell/\mu)}{\ell/\mu}\Big]_+ \bigg\} \,.
\end{align}
We see explicitly that large logs in $S_{\rm part}(\ell-\tilde\ell,\mu)$ are
minimized for $\mu\sim \ell-\tilde \ell$.  Hence when $\ell$ and $\tilde\ell$
are parametrically different it is the larger of the two that is important for
the proper setting of the renormalization scale in the soft function. This is
compatible with the expansion in Eq.~(\ref{Sop}). In the convolution with the
jet functions in the tail region in Eq.~(\ref{SFactThm}), the logs in $S_{\rm
  part}$ are minimized for $\mu=\mu_{op}$, and $S_{\rm part}(\ell^\pm,\mu_{op})$
can be determined by a truncated series in $\alpha_s(\mu_{op})$.  Thus for
Eq.~(\ref{SFactThm}) the result in Eq.~(\ref{S1}) works at any order in
perturbation theory.

We would also like $S(\ell^\pm,\mu)$ to give a viable model for the peak region
$S_{\rm np}(\ell^\pm,\mu)$ in Eq.~(\ref{FactThm}) when it is applied at a low
scale $\mu=\mu_{\rm low}\gtrsim \Lambda_{\rm QCD}$.  Here $\ell^\pm \sim \tilde
\ell^\pm$ in $S_{\rm part}(\ell^\pm\!-\tilde \ell^\pm,\mu)$ for the convolution
in Eq.~(\ref{S1}). This convolution builds the proper $\mu$-dependence into
$S(\ell^\pm,\mu)$, since the $\mu$-dependence is determined by perturbation
theory exactly as in $S_{\rm part}(\ell^\pm,\mu)$. Thus it avoids the issue of
having a $\mu$-dependence related to the soft function anomalous dimension in
the model parameters in $S_{\rm mod}$. The convolution with $S_{\rm part}$ also
generates a perturbative tail, implying that $S(\ell^\pm,\mu)$ is not
normalizable. To see this define the cutoff moments
\begin{align} \label{SLnm}
 S^{L[n,m]} \equiv
  \int_{-\infty}^L\!\!\! d\ell^+ \!\! \int_{-\infty}^{L} \!\!\!\! d\ell^-\:
  (\ell^+)^n (\ell^-)^m S(\ell^+,\ell^-,\mu) \,.
\end{align}
Using Eq.~(\ref{S1}) with Eq.~(\ref{SNLO}) and $S_{\rm mod}^{[0,0]}=1$ one finds
that for $L\gg \Lambda_{\rm QCD}$ the normalization
\begin{align}
\label{Snorm}
  S^{L[0,0]} = 1 + \frac{C_F\alpha_s(\mu)}{\pi}\bigg\{ \frac{\pi^2}{12}-
  2\ln^2\Big(\frac{L}{\mu}\Big) \bigg\} + \ldots \,,
\end{align}
up to terms of ${\cal O}(\alpha_s^2)$ or ${\cal O}(\Lambda_{\rm QCD}/L)$.
Rather than a deficiency, this behavior of $S^{L[0,0]}$ is a necessary
feature, as it is consistent with the renormalization equations for
$S(\ell^\pm,\mu)$. Only $S_{\rm mod}$ needs to be normalized.

For the peak region, perturbative improvements to $S_{\rm part}$ in
Eq.~(\ref{S1}) that cause a large change to $S$, could in principle be
compensated by changes to the model parameters in $S_{\rm mod}$. However, it is
quite desirable to make $S_{\rm part}$ and $S_{\rm mod}$ as independent as
possible, so that the interpretation of the model parameters remains unchanged
as we perturbatively improve $S_{\rm part}$.  A measure for this independence is
the convergence of the perturbative expansion for $S_{\rm part}$ at $\mu_{\rm
  low}$.  In general the convolution in Eq.~(\ref{S1}) generates double
logarithmic terms, $\ln^2(\ell/\mu_{\rm low})\sim\ln^2(\Lambda_{\rm
  QCD}/\mu_{\rm low})$ in $S(\ell^\pm,\mu_{\rm low})$ where the scale
$\Lambda_{\rm QCD}$ is set by parameters in $S_{\rm mod}$.  The choice of
$\mu_{\rm low}$ should be small enough to avoid these potentially large
logarithms, but large enough to ensure the validity of the perturbative
expansion in $\alpha_s(\mu_{\rm low})$. Thus a satisfactory choice of $\mu_{\rm
  low}$ might be difficult to find, and requires careful examination.  To test
this issue we can determine the logarithmic series for $S_{\rm
  part}(\ell^\pm,\mu)$, by finding the partonic soft function in
renormalization group improved perturbation theory at LL order, NLL order, etc.
The renormalization group improved $S_{\rm part}$ satisfies the exact relation
\begin{align}\label{SUS}
  S_{\rm part}(\ell^+,\ell^-,\mu) &= \int\!\! d\ell^{\,\prime +} d\ell^{\,\prime
    -} \: U_s(\ell^+\minus \ell^{\,\prime +},\mu,\mu_0)U_s(\ell^-\minus
  \ell^{\,\prime -},\mu,\mu_0)\: S_{\rm part}(\ell^{\,\prime +},\ell^{\,\prime
    -},\mu_0) \,,
\end{align}
where $U_s$ is the LL, NLL, etc. evolution kernel. As indicated this kernel
factors in the variables $\ell^+$ and $\ell^-$ to any order in perturbation
theory~\cite{FHMS2}.  Using this RG-improved $S_{\rm part}(\ell^\pm,\mu)$ the
full $S(\ell^\pm,\mu)$ in Eq.~(\ref{S1}) also satisfies the evolution
equation~(\ref{SUS}) exactly, with a $\mu$-independent $S_{\rm mod}(\tilde
\ell^\pm)$. When the logs are small we can expand the RG-improved result to a
fixed order in $\alpha_s(\mu)$, and the resulting $S_{\rm part}$ and $S$ satisfy
the RG to this order.  We will use this truncated version of the NLL series for
$S_{\rm part}(\ell^\pm,\mu)$ to test for a choice of $\mu$ which minimizes large
logs in the soft function.  This will also provide a test for the stability of
model parameters to the addition of perturbative corrections.

Lets construct the NLL partonic soft function using a Fourier transform as
in~\cite{Korchemsky:1993uz}.  At NLL order the partonic soft functions factorize
$S_{\rm part}^{NLL}(\ell^+,\ell^-) = S_{\rm part}^{NLL}(\ell^+) S_{\rm
  part}^{NLL}(\ell^-)$.  The Fourier transform of $ S_{\rm part}(\ell) = \int
d\ell'\, U_s(\ell-\ell',\mu,\mu_0)S_{\rm part}(\ell',\mu_0)$ is a simple product
equation
\begin{align} \label{tSUS}
  \tilde S_{\rm part}(y,\mu) 
    = \tilde U_s(y,\mu,\mu_0)\:\tilde S_{\rm part}(y,\mu_0) \,,
\end{align}
where the position space kernel is
\begin{align} 
\tilde U(y,\mu,\mu_0) = \big( i\, y\, \mu_0 e^{\gamma_E} \big)^{\omega(\mu,\mu_0)}
   \: e^{K(\mu,\mu_0)} 
\,.
\end{align}
The LL results for $\omega$ and
$K$ involve $\Gamma_0^{\rm cusp}$ and $\beta_0$ and the NLL results involve
$\Gamma_1^{\rm cusp}$ and $\beta_1$,
\begin{align} \label{wLS}
  \omega(\mu,\mu_0) &= 
     \frac{\Gamma_0^{\rm cusp}}{\beta_0}\, \bigg[ \ln(r) 
   + \bigg(\frac{\Gamma_1^{\rm cusp}}{\Gamma_0^{\rm cusp}}
     -\frac{\beta_1}{\beta_0}\bigg) \frac{\alpha_s(\mu_0)}{4\pi}\, (r-1) \bigg]
     \,,\\
  K(\mu,\mu_0) &= \frac{2\pi \Gamma_0^{\rm cusp}}{\beta_0^2}\:
  \bigg\{ \frac{1}{ \alpha_s(\mu)} \big( r\minus 1\minus r\ln r \big)  + \bigg(\frac{\Gamma_1^{\rm cusp}}{\Gamma_0^{\rm cusp}}
     -\frac{\beta_1}{\beta_0}\bigg) \frac{(1\minus r\plus \ln r)}{4\pi}
     +\frac{\beta_1}{8\pi\beta_0} \ln^2 r \bigg\}
   \,, \nn
\end{align}
which also agrees with Ref.~\cite{Neubert:2004dd}. Here
$r=\alpha_s(\mu)/\alpha_s(\mu_0)$, $C_F=4/3$, $\beta_0=11-2/3 n_f$ and $\beta_1=
34 C_A^2/3-10 C_A n_f/3 - 2 C_F n_f$ for $n_f$ light flavors, and the one and
two-loop terms of the cusp-anomalous dimension are $\Gamma_0=4C_F$ and
$\Gamma_1^{\rm cusp}= 4 C_F [ (67/9-\pi^2/3)C_A-10
n_f/9]$~\cite{Korchemsky:1987wg}.  To obtain a suitable boundary condition to
solve Eq.~(\ref{tSUS}) exactly, we note that the series of
$[\theta(\ell)\ln^k(\ell/\mu)/\ell]_+$ plus-functions in $S_{\rm
  part}(\ell,\mu)$ become a series of $\ln^k[i\, y\, \mu e^{\gamma_E}]$ in
$\tilde S_{\rm part}(y,\mu)$. Thus in position space we can take a boundary
condition where all the logs are absent. For example, using the LO boundary
condition $\tilde S_{\rm part}(y,\mu=-ie^{-\gamma_E}/y)=1$ in Eq.~(\ref{tSUS})
we obtain $\tilde S^{NLL}_{\rm part}(y,\mu)=\exp[K(\mu,-ie^{-\gamma_E}/y)]$. It
is straightforward to verify that this partonic soft function satisfies the
evolution equation in Eq.~(\ref{tSUS}).  Specifying higher order boundary
conditions for $\tilde S_{\rm part}$ will then properly specify the subleading
non-log terms in the series for $\tilde S_{\rm part}$.  For instance, $\tilde
S_{\rm part}(y,\mu=-ie^{-\gamma_E}/y) = 1 -\pi C_F \alpha_s(-i
e^{-\gamma_E}/y\mu)/8$ fixes the NLO boundary condition of Eq.~(\ref{SNLO}).
Thus the general solution to Eq.~(\ref{SUS}) is
\begin{align}
 S_{\rm part}(\ell,\mu) = \int \frac{dy}{2\pi}\: e^{i\, y\, \ell}\: 
   \:  \tilde S_{\rm part}(y,-ie^{-\gamma_E}/y)\: \exp\big[K(\mu,-i
   e^{-\gamma^E}/y)\big]
  \,.
\end{align}
This result allows us to determine the LL and NLL series. Order by order in
perturbation theory the Fourier transform (FT) can be carried out analytically
since
\begin{align}
  {\rm FT}\big[ \ln^k(i\, y\, \mu e^{\gamma_E} )\big] = 
  \frac{d^k}{d\epsilon^k} \: \frac{e^{\epsilon\gamma_E}}{\Gamma(1\minus\epsilon)}
  \bigg\{ \delta(\ell)  -
    \frac{\epsilon}{\mu}\,
    \left[\frac{\theta(\ell) e^{-\epsilon \ln(\ell/\mu)}}{\ell/\mu}\right]_+
  \bigg\} \bigg|_{\epsilon=0}
  \,.
\end{align}
In addition to the leading logs, this inverse Fourier transform gives
contributions to non-log terms from the expansion of
$e^{\epsilon\gamma_E}/\Gamma(1-\epsilon)$, which are subleading to the momentum
space NLL series. As long as such subleading terms are unambiguously defined order
by order and obey the RGE, one is free to include them in the NLL result. For our
purposes we define the LL, NLL, etc.~results as the resummed series obtained in
position space, since it is in this space that the evolution equations are the
simplest.  With the NLO boundary condition and NLL evolution we find
\begin{align}
\label{SLLexp}
  & S_{\rm part}(\ell,\mu) = \delta(\ell) + \frac{\alpha_s(\mu) C_F}{\pi} \Big[ 
   \minus 2 {\cal L}^1 \plus \frac{\pi^2}{24} \,\delta(\ell) \Big] 
  +
   \frac{\alpha_s^2(\mu)}{\pi^2} \bigg[ 
   C_F^2 \Big\{ 2 {\cal L}^3 \minus \frac{3\pi^2}{4} {\cal L}^1\plus
   4 \zeta_3 {\cal L}^0 \minus \frac{\pi^4}{80} \delta(\ell) \Big\}
 \nn\\
 & \qquad 
  + C_F \beta_0 \Big\{ \frac{{\cal L}^2}{2} - \frac{\pi^2}{48}\,{\cal L}^0
   + \frac{\zeta_3}{3} \delta(\ell) \Big\} 
 - \Gamma_1^{\rm cusp} \Big\{\frac{{\cal L}^1}{8} -
    \frac{\pi^2}{96}\delta(\ell) \Big\}
  \bigg] 
 \nn\\[0pt]
 &\
 + \frac{\alpha_s^3(\mu)}{\pi^3} \bigg[ C_F^3 \Big\{ \minus{\cal L}^5 \plus
  \frac{17\pi^2}{12} {\cal L}^3 \minus 20\zeta_3 {\cal L}^2 \plus
   \frac{\pi^4}{24} {\cal L}^1 \plus
   \Big(\frac{17\pi^2\zeta_3}{6} \minus 24\zeta_5 \Big) {\cal L}^0 
  \plus \Big(\frac{79\pi^6}{20160}\minus \frac{20\zeta_3^2}{3}\Big)\delta(\ell)\Big\} 
  \nn\\ 
 &\qquad + C_F^2\beta_0 \Big\{ \minus \frac{5{\cal L}^4}{6} \plus \frac{7\pi^2}{12}
 {\cal L}^2 -\frac{20\zeta_3}{3} {\cal L}^1 \plus \frac{\pi^4}{36} {\cal L}^0
 \plus \Big(\frac{7\pi^2\zeta_3}{18} \minus 4\zeta_5\Big)\delta(\ell) \Big\}
 \nn\\
 &\qquad + C_F \beta_0^2 \Big\{ \minus \frac{ {\cal L}^3}{6} \plus
 \frac{\pi^2}{48} {\cal
   L}^1 \minus \frac{\zeta_3}{3} {\cal L}^0 \plus \frac{13\pi^4}{2880}
 \delta(\ell)\Big\}
  + C_F \Gamma_1^{\rm cusp} \Big\{ \frac{{\cal L}^3}{4} \minus
 \frac{7\pi^2}{64} {\cal L}^1 \plus \frac{\zeta_3}{2} {\cal
   L}^0\minus \frac{\pi^4\delta(\ell)}{3840}\Big\} 
  \nn\\
 &\qquad + \Big( C_F \beta_1 + \frac12 \beta_0 \Gamma_1^{\rm cusp} \Big)
  \Big\{ \frac{{\cal L}^2}{8}  -\frac{\pi^2}{48} {\cal L}^0 +
 \frac{\zeta_3}{12} \delta(\ell)\Big\}
 \bigg] +   {\cal O}(\alpha_s^4)\,,
\end{align}
where ${\cal L}^j= 1/\mu \big[\theta(\ell)\ln^j(\ell/\mu)/(\ell/\mu)\big]_+$.
Note that the coefficients for the terms beyond NLL order are incomplete, namely
$\alpha_s^2 {\cal L}^0$, $\alpha_s^3 {\cal L}^{2,1,0}$, and
$\alpha_s^{2,3}\delta(\ell)$.  We show coefficients for these terms because of
our convention of specifying the series in position space and using the full
transform to momentum space. To obtain the complete $\alpha_s^2 {\cal L}^0$ and
$\alpha_s^3 {\cal L}^2$ terms we would need to include the non-cusp part of the
two-loop anomalous dimension.

Having determined the desired form of $S_{\rm part}$ in Eq.~(\ref{S1}) and a
means to test for large logs, we now turn to the nonperturbative information in
$S_{\rm mod}$ and the overlap with perturbation theory. To satisfy the moment
constraints on $S_{\rm mod}^{[n,m]}$ one can consider a two parameter model with
exponential tails~\cite{Korchemsky:2000kp}
\begin{align} \label{SM1}
  f_{\rm exp}(\tilde\ell^+,\tilde\ell^-) = \theta(\tilde\ell^+)\theta(\tilde\ell^-)
   \frac{ {\cal N}(a,b) }{\Lambda^2}
  \Big( \frac{\tilde\ell^+\tilde\ell^-}{\Lambda^2}\Big)^{a-1} \exp\Big(
  \frac{-(\tilde\ell^+)^2-(\tilde\ell^-)^2-2 b
    \tilde\ell^+\tilde\ell^-}{\Lambda^2} \Big)
\,,
\end{align}
where ${\cal N}(a,b)$ ensures $f_{\rm exp}$ is normalized to one, and $b\neq 0$
controls the noninclusive correlation between $\tilde\ell^+$ and $\tilde\ell^-$.
Physically the range $-1 < b < 0$ is favored~\cite{Korchemsky:2000kp}.  In the
past this and other models used for soft functions in jet physics are taken to
be nonzero for $\tilde\ell^\pm\ge 0$. This is a natural constraint given that it
is satisfied to any order in perturbation theory for $S_{\rm
  part}(\tilde\ell^\pm)$. With $\tilde \ell^\pm\ge 0$, Eq.~(\ref{S1}) enforces
$\ell^\pm\ge 0$ in $S(\ell^\pm,\mu)$.  However, a better approximation is to
take a soft-function with a gap so that the soft-function model vanishes for $\tilde
\ell^\pm < \Delta$,
\begin{align} \label{Sgap}
  S_{\rm mod}(\tilde\ell^+,\tilde\ell^-)
    =  
  f_{\rm exp}(\tilde\ell^+-\Delta,\tilde\ell^- - \Delta) \,.
\end{align}
Here $\Delta$ encodes the minimum hadronic energy deposit in each
hemisphere.\footnote{An even more accurate description of the gap would use
  $\ell^+\ell^-\ge m_{X_{\rm min}}^2$, but here there is a $\ell^\pm$ beyond
  which $S_{\rm mod}$ is exponentially suppressed, so the difference to
  Eq.~(\ref{Sgap}) is very small.}  Since the model parameter $\Delta\sim
\Lambda_{\rm QCD}$ it has an ${\cal O}(1)$ effect in the tail region where the
soft function is nonperturbative. Among the model parameters $\Delta$
plays a special role 
because it enables a hadronic interpretation for the variables $\tilde
\ell^\pm\ge \Delta$ in $S_{\rm mod}(\tilde\ell^\pm)$. 

Through the convolution in Eq.~(\ref{S1}) this gap is transferred to give
$\ell^\pm\ge \Delta$ in $S(\ell^\pm,\mu)$. This transfer relies on the fact that
we have a partonic threshold at zero-momentum, i.e.~that $S_{\rm part}(\ell^\pm
- \tilde\ell^\pm)$ has support only for $\ell^\pm\ge \tilde \ell^\pm$.
However, this transfer is not entirely straightforward because in perturbation
theory the partonic threshold has a renormalon which yields an ${\cal
  O}(\Lambda_{\rm QCD})$ ambiguity in $\Delta$. In the Borel transform of the
hemisphere soft function considered here, this renormalon corresponds to a pole
at $u=1/2$. Since the soft function is universal for massless jets and top quark
jets this renormalon is also behind the $u=1/2$ Borel pole identified by
Gardi~\cite{Gardi:2000lr} in an analysis of event-shape distributions in full
QCD for massless partons. The nature of this soft function renormalon is similar
to the well known ${\cal O}(\Lambda_{\rm QCD})$ renormalon of the heavy quark
pole mass definition, but is not equivalent to it; rather it is specific to the
soft function for jets.  For example, the $u=1/2$ renormalon pole of the soft
function that occurs in inclusive $B$ decays is solely related to the heavy
quark pole mass, and is eliminated by switching to a short-distance threshold
mass, see for example~\cite{Bosch:2004th}. For the case of the top jet event
shape distribution considered here the pole mass renormalon is contained in the
jet functions~\cite{Fleming:2007qr}, and is of no concern for the construction
of the soft function. Only gluon fields appear in the matrix element defining
our $S$ in Eq.~(\ref{Shemi}).

Using standard renormalon calculus either based on gluon propagators dressed
with massless fermion bubble chains or on the modified gluon propagator
\begin{align}
\label{gluonmod}
\frac{1}{q^2+i 0}\, \to \,
\left(\frac{e^{5/3}}{\mu^2}\right)^{-u}\frac{-1}{(-q^2-i 0)^{1+u}}
 \,,
\end{align}
the one-gluon exchange graphs in Fig.~\ref{softgraphs} give the Borel transform
\begin{align}
\label{borelStree}
B\Big[S_{\rm part}^{\rm
tree}(\ell^+,\ell^-,\mu)\Big]\Big(u\approx\frac{1}{2}\Big)
\,=\,
\frac{8 C_F e^{-5/6}}{\pi \beta_0\,(u-\frac{1}{2})}\,\mu\,
\Big(\,\delta(\ell^+)\delta^\prime(\ell^-)+
\delta^\prime(\ell^+)\delta(\ell^-)\,\Big)
\,.
\end{align}
This parameterizes the leading ${\cal O}(\Lambda_{\rm QCD})$ renormalon
ambiguity of the tree-level soft function, and has the same form as a shift in
the zero point of $\delta(\ell^+)\delta(\ell^-)$ expanded to first order. It is
also consistent with the result found by Gardi~\cite{Gardi:2000lr} for thrust,
accounting for Eq.~(\ref{Sthrust}).  Equation~(\ref{borelStree}) can be
generalized to soft function diagrams with an arbitrary number of gluons with
one gluon modified by Eq.~(\ref{gluonmod}).  Since one can use the soft limit
for the modified gluon momentum (compared to the momenta of the unmodified
gluons) only diagrams where the dressed gluon is external need to be considered.
The computation of the contributions from the dressed gluon then factorizes from
the remaining gluons yielding
\begin{align}
\label{borelS}
B\Big[S_{\rm part}(\ell^+,\ell^-,\mu)\Big]\Big(u\approx\frac{1}{2}\Big)
\,=\,
\frac{8 C_F e^{-5/6}}{\pi \beta_0\,(u-\frac{1}{2})}\,\mu\,
\Big(\,\frac{\partial}{\partial \ell^+}+
\frac{\partial}{\partial \ell^-}\,\Big)S_{\rm part}(\ell^+,\ell^-,\mu)
\,.
\end{align}
This result parameterizes the leading ${\cal O}(\Lambda_{\rm QCD})$ renormalon
ambiguity of the soft function at any order.  The Borel pole at $u=1/2$ leads to
instabilities in the perturbative predictions as we systematically include
perturbative corrections to $S_{\rm part}$. As we will see below, such
instabilities are for example reflected in $S$ becoming negative in certain
ranges of $\ell^\pm$, or in an instability of the $\ell^\pm$ values where $S$ is
maximal. Physically, this ambiguity ties together the perturbative physics that
we aimed to associate with $S_{\rm part}$ and the hadronic information in
$S_{\rm mod}$, and it must be resolved by experimental information.

In order to remove the ambiguity and allow for a stable determination
from experimental data we would like to use a renormalon free scheme
for the gap. Thus we take $\Delta = \bar \Delta + \delta$ where $\bar \Delta$ is
a renormalon-free model parameter for the hadronic threshold, and $\delta =
\delta_1 + \delta_2+\ldots $ has a perturbative expansion which cancels the
renormalon ambiguity in $S_{\rm part}$. Shifting variables to $\bar \ell^\pm =
\tilde \ell^\pm -\delta$ we have
\begin{align}
\label{S2}
S(\ell^+,\ell^-,\mu) & = 
\int_{-\infty}^{+\infty}\!\!\! d\bar\ell^+
\int_{-\infty}^{+\infty}\!\!\! d\bar\ell^-\
S_{\rm part}(\ell^+ \minus \bar\ell^+ \minus \delta,
  \ell^- \minus \bar\ell^-\minus \delta,\mu)\,
S_{\rm mod}(\bar\ell^+ \plus \delta,\bar\ell^- \plus \delta)
\nn\\
 & = 
\int_{-\infty}^{+\infty}\!\!\! d\bar\ell^+
\int_{-\infty}^{+\infty}\!\!\! d\bar\ell^-\
S_{\rm part}(\ell^+ \minus \bar\ell^+ \minus \delta,
  \ell^- \minus \bar\ell^-\minus \delta,\mu)\,
f_{\rm exp}(\bar\ell^+ \minus \bar\Delta,\bar\ell^- \minus \bar\Delta) \,.
\end{align}
To cancel the renormalon ambiguity we must expand Eq.~(\ref{S2}) in $\delta$
simultaneously with our expansion for $S_{\rm part}=S_{\rm part}^0+S_{\rm
  part}^1+\ldots$, so that
\begin{align}
\label{Sexpand}
  & S_{\rm part}(\ell^\pm \minus \delta,\mu)
  = S_{\rm part}^0(\ell^\pm,\mu) + \bigg[ S_{\rm part}^1(\ell^\pm,\mu) -\delta_1 
   \Big(\frac{d}{d\ell^+}\plus \frac{d}{d\ell^-}\Big) S_{\rm part}^0(\ell^\pm,\mu)
   \bigg]\nn\\
 &\qquad + \bigg[ S_{\rm part}^2(\ell^\pm,\mu) -
   \Big(\frac{d}{d\ell^+}\plus \frac{d}{d\ell^-}\Big) \Big\{ \delta_2 S_{\rm part}^0(\ell^\pm,\mu) +\delta_1 S_{\rm part}^1(\ell^\pm,\mu) \Big\}\nn\\
 &\qquad\quad + 
   \Big(\frac{d^2}{d\ell^{+\,2}}\plus \frac{d^2}{d\ell^{-\,2}} \plus 
   2 \frac{d^2}{d\ell^+d\ell^-}\Big) \frac{\delta_1^{\, 2}}{2} 
   S_{\rm part}^0(\ell^\pm,\mu)
   \bigg]
 + \ldots\,.
\end{align}
Here $\delta_i\sim {\cal O}(\alpha_s^i)$ can be defined with any prescription
that removes the ${\cal O}(\Lambda_{\rm QCD})$ renormalon ambiguity, and
simultaneously this prescription will define a scheme for the hadronic parameter
$\bar \Delta$.  Note that $\Delta$ is renormalization group invariant, thus
$\bar\Delta$ inherits a scale-dependence if $\delta$ is not renormalization
group invariant.  Moreover, we note that quadratic and higher powers of
$\delta_i$ that appear in Eq.~(\ref{Sexpand}) are required to ensure the
consistency of the perturbative scheme. The terms linear in $\delta_i$ are
the ones relevant for removing the leading ${\cal O}(\Lambda_{\rm QCD})$
ambiguity, having the same form as Eq.~(\ref{borelS}).

In order to motivate a definition for a subtraction scheme associated to
$\delta$ consider the first moment $S^{L[1,0]}$ from Eq.~(\ref{SLnm}).  For now
the upper cutoff $L$ is arbitrary.  Starting from Eq.~(\ref{S2}) we use the OPE
as in Eq.~(\ref{Sop}) and expand to linear order in $\delta$ to obtain
\begin{align} \label{SL10}
  S^{L[1,0]} &= S^{L[1,0]}_{\rm part} - 
\big[ S_{\rm mod}^{[1,0]}(\bar\Delta)+\delta \big]
\int_{-\infty}^L\!\!\! d\ell^+ \!\! \int_{-\infty}^{L} \!\!\!\! d\ell^-\:
 \ell^+
\left[\frac{\partial}{\partial\ell^+}+\frac{\partial}{\partial\ell^-}\right]
S_{\rm part}(\ell^+,\ell^-,\mu) 
 \nn\\
 &= S^{L[1,0]}_{\rm part} - 
\delta
\int_{-\infty}^L\!\!\! d\ell^+ \!\! \int_{-\infty}^{L} \!\!\!\! d\ell^-\:
 \ell^+
\left[\frac{\partial}{\partial\ell^+}+\frac{\partial}{\partial\ell^-}\right]
S_{\rm part}(\ell^+,\ell^-,\mu) 
 + S_{\rm mod}^{[1,0]}(\bar\Delta)
\,,
\end{align} 
where in the second line we dropped $\alpha_s$ corrections to the power
correction, and here
\begin{align} \label{SL10mod}
S_{\rm mod}^{[1,0]}(\bar\Delta) &
\, = \,
\int_{-\infty}^{+\infty}\!\!\! d\tilde\ell^+ \!\!\int_{-\infty}^{+\infty}\!\!\! 
d\tilde\ell^-\:
\tilde\ell^+\,f_{\rm exp}(\tilde\ell^+-\bar\Delta,\tilde\ell^--\bar\Delta)
\nn \\ &
\, = \,
\bar\Delta + 
\int_{-\infty}^{+\infty}\!\!\! d\tilde\ell^+ \!\!
\int_{-\infty}^{+\infty}\!\!\! d\tilde\ell^-\:
\tilde\ell^+\,f_{\rm exp}(\tilde\ell^+,\tilde\ell^-)
\,.
\end{align}
When $S_{\rm part}$ in the factorization theorem in Eq.~(\ref{SFactThm}) is
replaced by the full soft function $S$, the moment $S^{L[1,0]}$ appears in the
small $\ell^\pm$ region, and relates the small momentum contribution in the
leading order factorization theorem with the first power correction. From
Eq.~(\ref{borelS}) it is clear that there is a ${\cal O}(\Lambda_{\rm QCD})$
renormalon ambiguity in $S_{\rm part}^{L[1,0]}$ which should be canceled by the
$\delta$--term in Eq.~(\ref{SL10}). A suitable form for $\delta$ to
render the leading order factorization theorem and the first power
correction renormalon free is
\begin{align}
\label{deltadef}
\delta \, = \,
\frac{ 
\int\limits_{-\infty}^L\!\! d\ell^+ \!\! 
\int\limits_{-\infty}^{L} \!\! d\ell^-\:
\ell^+\,S_{\rm part}(\ell^+,\ell^-,\mu)
}{
\int\limits_{-\infty}^L\!\! d\ell^+ \!\! 
\int\limits_{-\infty}^{L} \!\! d\ell^-\:
\ell^+\,
{\displaystyle
\left[\frac{\partial}{\partial\ell^+}+
  \frac{\partial}{\partial\ell^-}\right]}\,
S_{\rm part}(\ell^+,\ell^-,\mu)
}\,. 
\end{align}  
Note that different choices of $L$ correspond to different schemes for
renormalon-free gap parameters $\bar\Delta$. Other ways to define $\delta$ are
also feasible.  From the expression for $S_{\rm part}$ given in
Eq.~(\ref{SLLexp}) we obtain
\begin{align}
\label{deltaexp}
\delta_1 \, = \, &
-2 L\,\frac{C_F\,\alpha_s(\mu)}{\pi}\,\left[\, \ln\frac{\mu}{L}  + 1 \right]
  \,,\nn\\
\delta_2 \, = \, & -\,L\,\frac{\alpha_s^2(\mu)}{\pi^2}\,\bigg\{
\beta_0\,C_F\,\left[\, \frac{1}{2}\ln^2\frac{\mu}{L} 
  + \ln \frac{\mu}{L} + 1 -\frac{\pi^2}{48}
  \right] + \Gamma_1^{\rm cusp}\left[ \frac{1}{8} \ln\frac{\mu}{L}  + \frac{1}{8}\right]
\nn\\ & \hspace{1.5cm}
 + C_F^2\,\left[ \Big(\frac{2\pi^2}{3}-8
   \Big) \ln\frac{\mu}{L}  + 4\zeta(3)+\frac{2\pi^2}{3}-12
  \right]
\bigg\} 
\,.
\end{align}
Note that the one-loop $\delta_1$ term is exact, while the two-loop term
$\delta_2$ relies on our NLL approximation of Eq.~(\ref{SLLexp}).

Lets examine the impact of renormalon subtractions on the soft function.
\begin{figure}[t]
\begin{center}
\epsfxsize=\textwidth
\epsffile{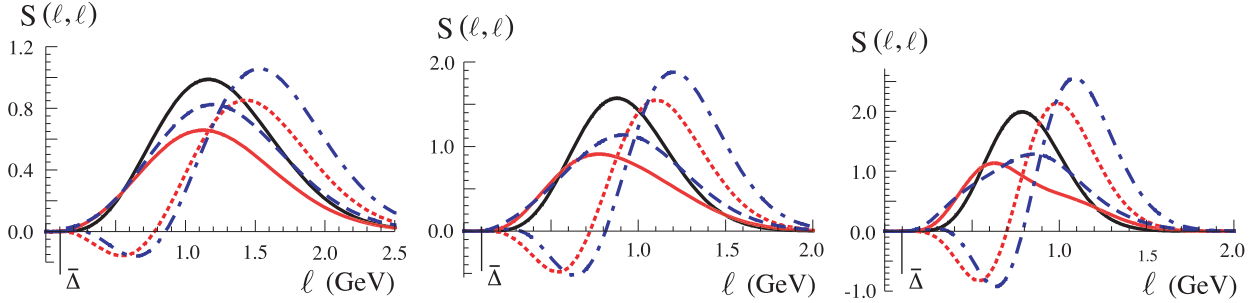}
\vspace{-1.3 cm}
\vskip 0.4cm
\caption{Soft function $S(\ell^+,\ell^-\mu)$ as a function of
  $\ell=\ell^+=\ell^-$ with $\mu=1\,{\rm GeV}$, at tree level (solid black
  line), one-loop (dotted red line), one-loop with renormalon subtraction (light
  solid red line), two-loop NLL (dot-dashed blue line), and two-loop NLL with
  renormalon subtraction (dashed blue line). Results are shown for three models:
  $(a,b)=(2.5,-0.8)$ (left panel), $(3.0,-0.5)$ (middle panel) and $(3.5,-0.2)$
  (right panel). All models have $\Lambda=0.55$~GeV and a gap of $\bar\Delta=100\,{\rm
    MeV}$.  }
\label{fig:softfct}
\end{center}
\end{figure}
In Fig.~\ref{fig:softfct} $S(\ell^+,\ell^-,\mu)$ is plotted as a function of
$\ell=\ell^+=\ell^-$ at tree-level (solid black line) and one-loop (dotted and
lighter solid red lines).  Blue dashed and dot-dashed lines are two-loop NLL
results to be discussed below. We take $\mu=1.0$~GeV ($\alpha_s(\mu)=0.396$) and
use the soft model function of Eq.~(\ref{SM1}) with $\Lambda=0.55$~GeV, and
three different choices $(a,b)=(2.5,-0.8)$ (left panel), $(3.0,-0.5)$ (middle
panel), and $(3.5,-0.2)$ (right panel). The dotted red line is the one-loop
corrected soft function prior to renormalon subtractions, with $\delta_1=0$ and
$\bar\Delta=\Delta$. The light solid red line is the corresponding result with a
renormalon free gap parameter $\bar\Delta$, and subtraction using $\delta_1$
from Eq.~(\ref{deltaexp}). We use $L=\Lambda$ as a representative scheme choice,
and for simplicity have chosen $\bar\Delta=100\,{\rm MeV}$.  Other values of
$\bar\Delta$ simply correspond to a global horizontal shift of all curves by the
same amount.  While the unsubtracted one-loop soft functions have unphysical
negative values for small $\ell$, we see that the renormalon-subtracted curves
are alway positive.  This effect of the renormalon subtraction is very general,
we have checked that it is realized for any choice of model parameters,
renormalization scale $\mu$, and scheme parameter $L\gsim\Lambda$. We illustrate
this in Fig.~\ref{fig:softfct2} by showing soft functions $S(\ell,\ell,\mu)$
with $\Lambda=0.55$~GeV and $(a,b)=(3,-0.5)$, for different choices of $\mu$ and
$L$. For the upper (lower) panels $\mu=1.0$ $(1.3)$~GeV, and for the left,
middle and right panels we have $L/\Lambda=0.5,1.0$ and $1.5$. Note that the
soft function has an anomalous dimension, see Eq.~(\ref{SUS}) and (\ref{Snorm}),
so its shape and normalization change when varying $\mu$.

In Fig.~\ref{fig:softfct} the subtracted curves also show a somewhat smaller
correction to the $\ell$ value where their maximum is located than the
unsubtracted curves, but this effect is more dependent on the choice of
parameters, such as the $L$ value, see Fig.~\ref{fig:softfct2}.  At ${\cal
  O}(\alpha_s)$ the perturbative series for the peak position has not yet
approached its asymptotic behavior, but we expect the improvement in convergence
for the peak position of the soft function to become more pronounced when higher
order perturbative results for the soft function are considered.

\begin{figure}[t]
\begin{center}
\epsfxsize=\textwidth
\epsffile{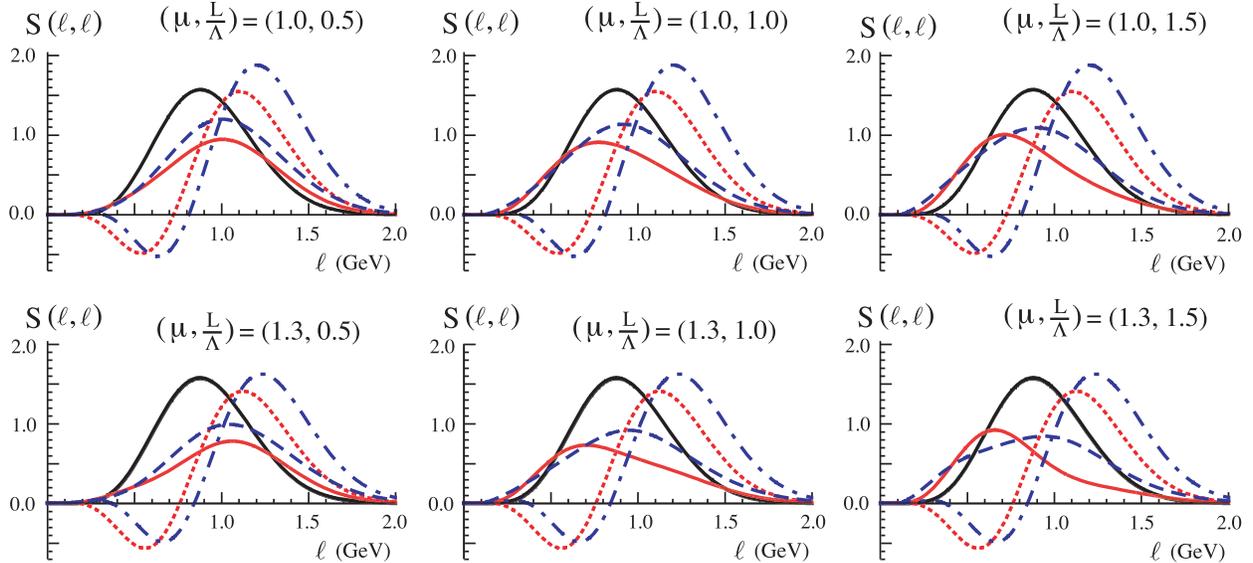}
\vspace{-1.3 cm}
\vskip 0.4cm
\caption{Dependence of the soft function $S(\ell,\ell,\mu)$ on the
  renormalization scale $\mu$ and the renormalon subtraction scheme-parameter
  $L$ for the model with $\Lambda=0.55$~GeV and $(a,b)=(3.0,-0.5)$. Lines use the
  same conventions as for Fig.~\ref{fig:softfct}. As indicated the upper and
  lower panels represent curves for $\mu=1.0$ and $1.3$~GeV, while
  the left, middle and right panels refer to $L/\Lambda=0.5$, $1.0$ and $1.5$.  }
\label{fig:softfct2}
\end{center}
\end{figure}

To test whether $S_{\rm part}$ suffers from large logs for particular values of
$\mu$, the $O(\alpha_s^2)$ NLL predictions for the soft function from
Eq.~(\ref{SLLexp}) are shown as the blue dot-dashed and dashed lines in
Figs.~\ref{fig:softfct} and \ref{fig:softfct2}.  The dot-dashed curves do not
have renormalon subtractions, and again exhibit negative dips. The dashed curve
use our renormalon free $\bar\Delta$, with subtractions given by the terms in the
last set of square brackets in Eq.~(\ref{Sexpand}) and $\delta_{1}$ and
$\delta_2$ from Eq.~(\ref{deltaexp}).  We see that at this order the renormalon
subtractions continue to eliminate the negative dip at small $\ell$ values.  The
behavior of the peak location for the two-loop NLL result is in general not
dramatically improved, but this is simply because the ${\cal O}(\alpha_s^2)$
soft function given in Eq.~(\ref{SLLexp}) is based on a logarithmic
approximation in a region where the logs are not large, and hence does not
contain the large renormalon terms of the full two-loop soft function.  Finally,
for the lower right panel of Fig.~\ref{fig:softfct2}, we see an indication for an
instability due to increasing logarithmic terms for $\mu=1.3$~GeV and
$L/\Lambda=1.5$. For the model function of Eq.~(\ref{SM1}) such regions of
instability generally arise for larger values of $\mu$, and increasing positive
values of $b$ and $L/\Lambda$. This issue might have to be more carefully
examined if experimental data suggests that such regions of model parameters are
favored.

The impact of the renormalon subtraction is also significant for the
differential cross section. Let us first consider the peak region based on the
factorization theorem~(\ref{FactThm}). Since we only wish to illustrate the
impact of the soft function, we use tree-level jet functions $B_\pm(\hat s)=
1/({\hat s}^2+\Gamma_t^2)$ for $Q/m_t=5$, $\Gamma_t=1.43$~GeV, $m_t=172$~GeV. We
also ignore common normalization factors, and evolution factors that sum
large logarithms down to the low renormalization scale $\mu_{\rm low}$ of the soft
function, since they affect all predictions in the same way.  (For the case of
top-quark jets, a complete analysis including all these terms is carried out in
Ref.~\cite{FHMS2}.)
\begin{figure}
\begin{center}
\epsfxsize=\textwidth
\epsffile{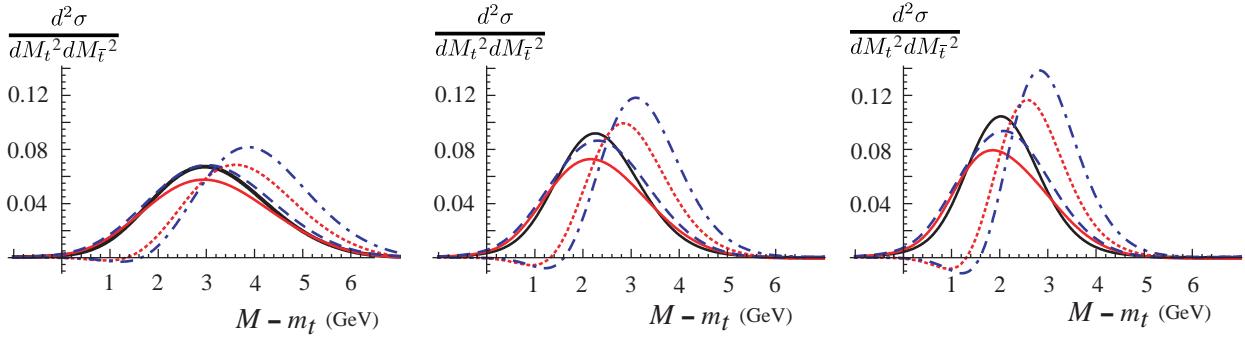}
\vspace{-1.3 cm}
\vskip 0.4cm
\caption{Top invariant mass distribution $d\sigma/dM_t^2dM_{\bar t}^2$ in
  the peak region as a function of $M-m_t$ with $M=M_t=M_{\bar t}$
  accounting only for the perturbative 
  corrections arising from the soft function. The left, middle and
  right panel refer to the respective models and renormalon
  subtraction scheme used in
  Fig.~\ref{fig:softfct} and the same line specifications are employed.
}
\label{fig:peak}
\end{center}
\end{figure}
Fig.~\ref{fig:peak} displays this differential cross section for equal invariant
masses $M=M_t=M_{\bar t}$ over $M-m_t$ for the three parameters sets of
Fig.~\ref{fig:softfct}. Again we find that using a renormalon free gap parameter
improves the convergence of the predictions and avoids the problem of negative
dips in the cross-section.  Interestingly, the curves show even better
convergence compared to the soft function alone, and show nice convergence for
the peak location. We find that this is true in general and related to the
additional smearing that is provided by the width of the jet function.  These
results illustrate that the removal of the ${\cal O}(\Lambda_{\rm QCD})$
renormalon contributions in the soft function is essential to obtain a
renormalon-free mass measurement from the peak position of the invariant mass
distribution. We emphasize again that the renormalon issue in the soft function
treated here is entirely independent of the pole mass renormalon problem, which
appears in the massive jet function and the top quark pole mass.

Finally, let us examine the tail region of the differential cross section, using
again tree-level jet functions and equal invariant masses $M=M_t=M_{\bar t}$ and
ignoring common normalization factors. To be specific we adopt the model with
$\Lambda=0.55$~GeV and $(a,b)=(3.0,-0.5)$.
\begin{figure}
\begin{center}
\epsfxsize=15cm
\epsffile{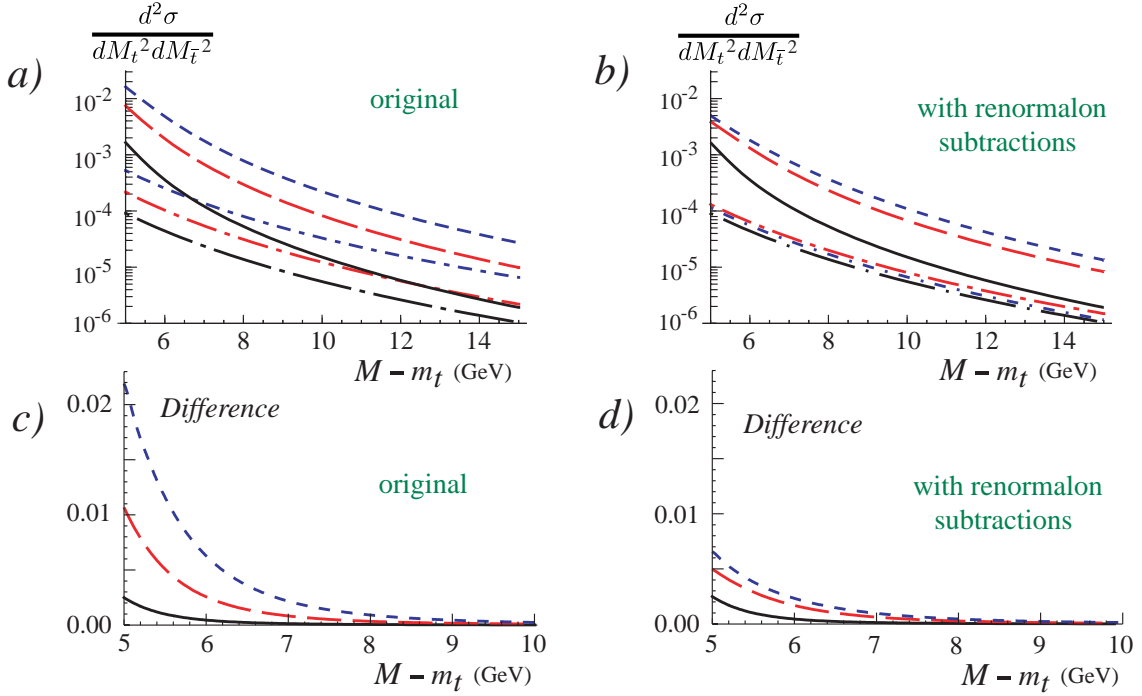}
\vskip -0.5cm
\caption{Top invariant mass distribution
  $d\sigma/dM_t^2dM_{\bar t}^2$ in the tail region as a function of $M-m_t$,
  with $M=M_t=M_{\bar t}$, $\mu=(M^2-m_t^2)/Q$, and $\bar\Delta=100\,{\rm MeV}$.
  In a) results are shown without renormalon subtraction ($\delta=0$), and in b)
  with a renormalon free gap parameter $\bar\Delta$.  In a),b) we show: three
  curves using $S_{\rm part}$ at tree, 1-loop, ${\cal O}(\alpha_s^2)$ NLL (long
  dot-dashed black, medium dot-dashed red, short dot-dashed blue), and three
  curvies using the full $S$ at tree, 1-loop, ${\cal O}(\alpha_s^2)$ NLL (solid
  black, long dashed red, short dashed blue). The latter three curves use the
  model with $\Lambda=0.55$~GeV and $(a,b)=(3.0,-0.5)$, and reflect the effects
  from power corrections when compared to the former three. In c) and d) we show
  the difference between the first and second set of three curves from a) and b)
  respectively, at tree (solid black), one-loop (long-dashed red), and ${\cal
    O}(\alpha_s^2)$ NLL (short-dashed blue). }
\label{fig:tail}
\end{center}
\end{figure}
In Fig.~\ref{fig:tail}a the tree-level (black lines), one-loop (red lines) and
two-loop (blue lines) cross sections are shown without renormalon subtractions
as a function of $M-m_t$. We use $\mu=(M^2-m_t^2)/Q$ to avoid large logs in the
soft function when plotting over a wide range of scales. The dot-dashed lines
use the leading order result in Eq.~(\ref{SFactThm}) with only the partonic soft
function and no gap, and the solid and dashed lines use the full soft function
$S$ from Eq.~(\ref{S1}) instead and take $\bar\Delta=100\,{\rm MeV}$.  For a
given order in $\alpha_s$ the difference between the curves in
Fig.~\ref{fig:tail}a reflect the typical size of power corrections, and are
plotted in Fig.~\ref{fig:tail}c. In Fig~\ref{fig:tail}b the same tail
distributions as Fig.~\ref{fig:tail}a are displayed, but now with the renormalon
subtraction.  Since the perturbative contributions in $S_{\rm part}$ are at the
scale $\mu_{\rm op}$ it is mandatory to choose $L$ of order $\mu_{\rm op}$ to
avoid large logarithmic terms, as can be also seen from Eq.~(\ref{deltadef}),
and we adopt the specific scheme choice $L=\mu_{\rm op}$.  Comparing the curves
in Figs.~\ref{fig:tail}a,b we see that the renormalon subtraction substantially
improves the perturbative convergence.  Figure~\ref{fig:tail}d shows the
difference between the solid/dashed and the dot-dashed curves from
Fig.~\ref{fig:tail}b.  Comparing it to Fig.~\ref{fig:tail}c we see that the
renormalon subtractions lead, as anticipated, to a significantly better
perturbative behavior for values one would extract from the data for the power
correction.\footnote{Note that our choice of a gap of $\bar\Delta=100\,{\rm
    MeV}$ shifts all curves in Fig.~\ref{fig:tail}a,b that use $S$ to larger values
  of $M-m_t$. This is a significant power correction, it increases these
  cross-sections by $\sim$30\%.  However, the choice of $\bar\Delta$ does not effect
  the impact of the renormalon subtraction.} This illustrates that the
renormalon subtracted predictions are essential for extracting stable and
renormalon-free model parameters from experimental data. A scheme such as the
one used here, where $L=\mu$, works well for both the tail and peak regions,
avoiding large logs. If a result for the gap model parameter is determined from
data in a scheme where $L=\mu$, then Eq.~(\ref{deltaexp}) can be used to relate
the result to other schemes, such as for $L=\mu/2$.

To conclude, we have provided a prescription for designing soft function models
in jet production, that can be applied both in the peak region where the soft
function is nonperturbative and in the tail region where the soft function can
be expanded with an OPE. The method entails the convolution of the partonic soft
function with a normalized model function that encodes the nonperturabive
information, Eq.~(\ref{S1}). It automatically implements consistent
renormalization scaling behavior in the $\overline{\rm MS}$ scheme, making the
design particularly useful when dimensional regularization is employed for
perturbative calculations.  As a novel feature we argue that the soft function
models need to exhibit a gap which accounts for the fact that for real hadrons
there is a minimal hadronic energy. This gap is also required to devise a
systematic scheme to remove the leading ${\cal O}(\Lambda_{\rm QCD})$ renormalon
that is contained in the partonic soft function. In
Eqs.~(\ref{S2},\ref{Sexpand}, \ref{deltadef}) we have provided a simple
definition for such a scheme and demonstrated that the removal of the renormalon
avoids large uncertainties in predictions of the soft function and hence the
cross-section in the peak region.  In the tail region it also reduces the size
of fluctuations in the power corrections, since they are otherwise affected by
the ${\cal O}(\Lambda_{\rm QCD})$ renormalon.  It is possible to generalize our
method to treat also subleading ${\cal O}(\Lambda_{\rm QCD}^n)$ renormalons with
$n>1$, which are expected to have smaller effect on the soft function stability.
Subtraction of these subleading renormalons might improve the numerical
stability at higher order in perturbation theory of model parameters in $S_{\rm
  mod}$ not related to the gap.

\vspace{-0.3cm}
 
\acknowledgments{ This work was supported in part by the Department of Energy
  Office of Nuclear Science under the grant DE-FG02-94ER40818, and in part by
  the EU network contract MRTN-CT-2006-035482 (FLAVIAnet). We thank the Aspen
  Center for Physics for the inspiring atmosphere provided during the workshop
  ``Between the LHC and B Factories'' where the bulk of this work was
  accomplished. We also thank S.~Fleming and S.~Mantry for their collaboration
  on related work~\cite{FHMS2}.  }

\newpage

\bibliography{topjet}

\begin{thebibliography}{29}
\expandafter\ifx\csname natexlab\endcsname\relax\def\natexlab#1{#1}\fi
\expandafter\ifx\csname bibnamefont\endcsname\relax
  \def\bibnamefont#1{#1}\fi
\expandafter\ifx\csname bibfnamefont\endcsname\relax
  \def\bibfnamefont#1{#1}\fi
\expandafter\ifx\csname citenamefont\endcsname\relax
  \def\citenamefont#1{#1}\fi
\expandafter\ifx\csname url\endcsname\relax
  \def\url#1{\texttt{#1}}\fi
\expandafter\ifx\csname urlprefix\endcsname\relax\def\urlprefix{URL }\fi
\providecommand{\bibinfo}[2]{#2}
\providecommand{\eprint}[2][]{\url{#2}}

\bibitem[{\citenamefont{Korchemsky}(1998)}]{Korchemsky:1998ev}
\bibinfo{author}{\bibfnamefont{G.~P.} \bibnamefont{Korchemsky}}
  (\bibinfo{year}{1998}),
  \eprint{\href{http://arXiv.org/abs/hep-ph/9806537}{hep-ph/9806537}},
  \urlprefix\url{http://arXiv.org/abs/hep-ph/9806537}.

\bibitem[{\citenamefont{Korchemsky and Sterman}(1999)}]{Korchemsky:1999kt}
\bibinfo{author}{\bibfnamefont{G.~P.} \bibnamefont{Korchemsky}}
  \bibnamefont{and} \bibinfo{author}{\bibfnamefont{G.}~\bibnamefont{Sterman}},
  \bibinfo{journal}{Nucl. Phys.} \textbf{\bibinfo{volume}{B555}},
  \bibinfo{pages}{335} (\bibinfo{year}{1999}),
  \eprint{\href{http://arXiv.org/abs/hep-ph/9902341}{hep-ph/9902341}},
  \urlprefix\url{http://arXiv.org/abs/hep-ph/9902341}.

\bibitem[{\citenamefont{Bauer et~al.}(2004)\citenamefont{Bauer, Lee, Manohar,
  and Wise}}]{Bauer:2003di}
\bibinfo{author}{\bibfnamefont{C.~W.} \bibnamefont{Bauer}},
  \bibinfo{author}{\bibfnamefont{C.}~\bibnamefont{Lee}},
  \bibinfo{author}{\bibfnamefont{A.~V.} \bibnamefont{Manohar}},
  \bibnamefont{and} \bibinfo{author}{\bibfnamefont{M.~B.} \bibnamefont{Wise}},
  \bibinfo{journal}{Phys. Rev.} \textbf{\bibinfo{volume}{D70}},
  \bibinfo{pages}{034014} (\bibinfo{year}{2004}),
  \eprint{\href{http://arXiv.org/abs/hep-ph/0309278}{hep-ph/0309278}},
  \urlprefix\url{http://arXiv.org/abs/hep-ph/0309278}.

\bibitem[{\citenamefont{Catani et~al.}(1991)\citenamefont{Catani, Turnock,
  Webber, and Trentadue}}]{Catani:1991kz}
\bibinfo{author}{\bibfnamefont{S.}~\bibnamefont{Catani}},
  \bibinfo{author}{\bibfnamefont{G.}~\bibnamefont{Turnock}},
  \bibinfo{author}{\bibfnamefont{B.~R.} \bibnamefont{Webber}},
  \bibnamefont{and}
  \bibinfo{author}{\bibfnamefont{L.}~\bibnamefont{Trentadue}},
  \bibinfo{journal}{Phys. Lett.} \textbf{\bibinfo{volume}{B263}},
  \bibinfo{pages}{491} (\bibinfo{year}{1991}).

\bibitem[{\citenamefont{Korchemsky and Sterman}(1995)}]{Korchemsky:1994is}
\bibinfo{author}{\bibfnamefont{G.~P.} \bibnamefont{Korchemsky}}
  \bibnamefont{and} \bibinfo{author}{\bibfnamefont{G.}~\bibnamefont{Sterman}},
  \bibinfo{journal}{Nucl. Phys.} \textbf{\bibinfo{volume}{B437}},
  \bibinfo{pages}{415} (\bibinfo{year}{1995}), \eprint{hep-ph/9411211}.

\bibitem[{\citenamefont{Dokshitzer and Webber}(1997)}]{Dokshitzer:1997ew}
\bibinfo{author}{\bibfnamefont{Y.~L.} \bibnamefont{Dokshitzer}}
  \bibnamefont{and} \bibinfo{author}{\bibfnamefont{B.~R.}
  \bibnamefont{Webber}}, \bibinfo{journal}{Phys. Lett.}
  \textbf{\bibinfo{volume}{B404}}, \bibinfo{pages}{321} (\bibinfo{year}{1997}),
  \eprint{hep-ph/9704298}.

\bibitem[{\citenamefont{Farhi}(1977)}]{Farhi:1977sg}
\bibinfo{author}{\bibfnamefont{E.}~\bibnamefont{Farhi}},
  \bibinfo{journal}{Phys. Rev. Lett.} \textbf{\bibinfo{volume}{39}},
  \bibinfo{pages}{1587} (\bibinfo{year}{1977}).

\bibitem[{\citenamefont{Catani et~al.}(1992)\citenamefont{Catani, Turnock, and
  Webber}}]{Catani:1992jc}
\bibinfo{author}{\bibfnamefont{S.}~\bibnamefont{Catani}},
  \bibinfo{author}{\bibfnamefont{G.}~\bibnamefont{Turnock}}, \bibnamefont{and}
  \bibinfo{author}{\bibfnamefont{B.~R.} \bibnamefont{Webber}},
  \bibinfo{journal}{Phys. Lett.} \textbf{\bibinfo{volume}{B295}},
  \bibinfo{pages}{269} (\bibinfo{year}{1992}).

\bibitem[{\citenamefont{Chandramohan and Clavelli}(1981)}]{Chandramohan:1980ry}
\bibinfo{author}{\bibfnamefont{T.}~\bibnamefont{Chandramohan}}
  \bibnamefont{and} \bibinfo{author}{\bibfnamefont{L.}~\bibnamefont{Clavelli}},
  \bibinfo{journal}{Nucl. Phys.} \textbf{\bibinfo{volume}{B184}},
  \bibinfo{pages}{365} (\bibinfo{year}{1981}).

\bibitem[{\citenamefont{Berger et~al.}(2003)\citenamefont{Berger, Kucs, and
  Sterman}}]{Berger:2002ig}
\bibinfo{author}{\bibfnamefont{C.~F.} \bibnamefont{Berger}},
  \bibinfo{author}{\bibfnamefont{T.}~\bibnamefont{Kucs}}, \bibnamefont{and}
  \bibinfo{author}{\bibfnamefont{G.}~\bibnamefont{Sterman}},
  \bibinfo{journal}{Int. J. Mod. Phys.} \textbf{\bibinfo{volume}{A18}},
  \bibinfo{pages}{4159} (\bibinfo{year}{2003}), \eprint{hep-ph/0212343}.

\bibitem[{\citenamefont{Fleming
  et~al.}(2007{\natexlab{a}})\citenamefont{Fleming, Hoang, Mantry, and
  Stewart}}]{Fleming:2007qr}
\bibinfo{author}{\bibfnamefont{S.}~\bibnamefont{Fleming}},
  \bibinfo{author}{\bibfnamefont{A.~H.} \bibnamefont{Hoang}},
  \bibinfo{author}{\bibfnamefont{S.}~\bibnamefont{Mantry}}, \bibnamefont{and}
  \bibinfo{author}{\bibfnamefont{I.~W.} \bibnamefont{Stewart}}
  (\bibinfo{year}{2007}{\natexlab{a}}), \eprint{hep-ph/0703207}.

\bibitem[{\citenamefont{Kidonakis et~al.}(1998)\citenamefont{Kidonakis, Oderda,
  and Sterman}}]{Kidonakis:1998ur}
\bibinfo{author}{\bibfnamefont{N.}~\bibnamefont{Kidonakis}},
  \bibinfo{author}{\bibfnamefont{G.}~\bibnamefont{Oderda}}, \bibnamefont{and}
  \bibinfo{author}{\bibfnamefont{G.}~\bibnamefont{Sterman}}
  (\bibinfo{year}{1998}), \eprint{hep-ph/9805279}.

\bibitem[{\citenamefont{Neubert}(1994)}]{Neubert:1994um}
\bibinfo{author}{\bibfnamefont{M.}~\bibnamefont{Neubert}},
  \bibinfo{journal}{Phys. Rev.} \textbf{\bibinfo{volume}{D49}},
  \bibinfo{pages}{4623} (\bibinfo{year}{1994}),
  \eprint{\href{http://arXiv.org/abs/hep-ph/9312311}{hep-ph/9312311}},
  \urlprefix\url{http://arXiv.org/abs/hep-ph/9312311}.

\bibitem[{\citenamefont{Bigi et~al.}(1994)\citenamefont{Bigi, Shifman,
  Uraltsev, and Vainshtein}}]{Bigi:1994ex}
\bibinfo{author}{\bibfnamefont{I.~I.~Y.} \bibnamefont{Bigi}},
  \bibinfo{author}{\bibfnamefont{M.~A.} \bibnamefont{Shifman}},
  \bibinfo{author}{\bibfnamefont{N.~G.} \bibnamefont{Uraltsev}},
  \bibnamefont{and} \bibinfo{author}{\bibfnamefont{A.~I.}
  \bibnamefont{Vainshtein}}, \bibinfo{journal}{Int. J. Mod. Phys.}
  \textbf{\bibinfo{volume}{A9}}, \bibinfo{pages}{2467} (\bibinfo{year}{1994}),
  \eprint{\href{http://arXiv.org/abs/hep-ph/9312359}{hep-ph/9312359}},
  \urlprefix\url{http://arXiv.org/abs/hep-ph/9312359}.

\bibitem[{\citenamefont{Mannel and Neubert}(1994)}]{Mannel:1994pm}
\bibinfo{author}{\bibfnamefont{T.}~\bibnamefont{Mannel}} \bibnamefont{and}
  \bibinfo{author}{\bibfnamefont{M.}~\bibnamefont{Neubert}},
  \bibinfo{journal}{Phys. Rev.} \textbf{\bibinfo{volume}{D50}},
  \bibinfo{pages}{2037} (\bibinfo{year}{1994}),
  \eprint{\href{http://arXiv.org/abs/hep-ph/9402288}{hep-ph/9402288}},
  \urlprefix\url{http://arXiv.org/abs/hep-ph/9402288}.

\bibitem[{\citenamefont{Lee and Stewart}(2006)}]{Lee:2005pk}
\bibinfo{author}{\bibfnamefont{K.~S.~M.} \bibnamefont{Lee}} \bibnamefont{and}
  \bibinfo{author}{\bibfnamefont{I.~W.} \bibnamefont{Stewart}},
  \bibinfo{journal}{Phys. Rev.} \textbf{\bibinfo{volume}{D74}},
  \bibinfo{pages}{014005} (\bibinfo{year}{2006}),
  \eprint{\href{http://arXiv.org/abs/hep-ph/0511334}{hep-ph/0511334}},
  \urlprefix\url{http://arXiv.org/abs/hep-ph/0511334}.

\bibitem[{\citenamefont{Lee et~al.}(2006)\citenamefont{Lee, Ligeti, Stewart,
  and Tackmann}}]{Lee:2005pw}
\bibinfo{author}{\bibfnamefont{K.~S.~M.} \bibnamefont{Lee}},
  \bibinfo{author}{\bibfnamefont{Z.}~\bibnamefont{Ligeti}},
  \bibinfo{author}{\bibfnamefont{I.~W.} \bibnamefont{Stewart}},
  \bibnamefont{and} \bibinfo{author}{\bibfnamefont{F.~J.}
  \bibnamefont{Tackmann}}, \bibinfo{journal}{Phys. Rev.}
  \textbf{\bibinfo{volume}{D74}}, \bibinfo{pages}{011501}
  (\bibinfo{year}{2006}), \eprint{hep-ph/0512191}.

\bibitem[{\citenamefont{Korchemsky and Tafat}(2000)}]{Korchemsky:2000kp}
\bibinfo{author}{\bibfnamefont{G.~P.} \bibnamefont{Korchemsky}}
  \bibnamefont{and} \bibinfo{author}{\bibfnamefont{S.}~\bibnamefont{Tafat}},
  \bibinfo{journal}{JHEP} \textbf{\bibinfo{volume}{10}}, \bibinfo{pages}{010}
  (\bibinfo{year}{2000}),
  \eprint{\href{http://arXiv.org/abs/hep-ph/0007005}{hep-ph/0007005}},
  \urlprefix\url{http://arXiv.org/abs/hep-ph/0007005}.

\bibitem[{\citenamefont{Bosch et~al.}(2004{\natexlab{a}})\citenamefont{Bosch,
  Lange, Neubert, and Paz}}]{Bosch:2004th}
\bibinfo{author}{\bibfnamefont{S.~W.} \bibnamefont{Bosch}},
  \bibinfo{author}{\bibfnamefont{B.~O.} \bibnamefont{Lange}},
  \bibinfo{author}{\bibfnamefont{M.}~\bibnamefont{Neubert}}, \bibnamefont{and}
  \bibinfo{author}{\bibfnamefont{G.}~\bibnamefont{Paz}},
  \bibinfo{journal}{Nucl. Phys.} \textbf{\bibinfo{volume}{B699}},
  \bibinfo{pages}{335} (\bibinfo{year}{2004}{\natexlab{a}}),
  \eprint{\href{http://arXiv.org/abs/hep-ph/0402094}{hep-ph/0402094}},
  \urlprefix\url{http://arXiv.org/abs/hep-ph/0402094}.

\bibitem[{\citenamefont{Bauer et~al.}(2002)\citenamefont{Bauer, Luke, and
  Mannel}}]{Bauer:2002yu}
\bibinfo{author}{\bibfnamefont{C.~W.} \bibnamefont{Bauer}},
  \bibinfo{author}{\bibfnamefont{M.}~\bibnamefont{Luke}}, \bibnamefont{and}
  \bibinfo{author}{\bibfnamefont{T.}~\bibnamefont{Mannel}},
  \bibinfo{journal}{Phys. Lett.} \textbf{\bibinfo{volume}{B543}},
  \bibinfo{pages}{261} (\bibinfo{year}{2002}), \eprint{hep-ph/0205150}.

\bibitem[{\citenamefont{Lee and Stewart}(2005)}]{Lee:2004ja}
\bibinfo{author}{\bibfnamefont{K.~S.~M.} \bibnamefont{Lee}} \bibnamefont{and}
  \bibinfo{author}{\bibfnamefont{I.~W.} \bibnamefont{Stewart}},
  \bibinfo{journal}{Nucl. Phys.} \textbf{\bibinfo{volume}{B721}},
  \bibinfo{pages}{325} (\bibinfo{year}{2005}),
  \eprint{\href{http://arXiv.org/abs/hep-ph/0409045}{hep-ph/0409045}},
  \urlprefix\url{http://arXiv.org/abs/hep-ph/0409045}.

\bibitem[{\citenamefont{Bosch et~al.}(2004{\natexlab{b}})\citenamefont{Bosch,
  Neubert, and Paz}}]{Bosch:2004cb}
\bibinfo{author}{\bibfnamefont{S.~W.} \bibnamefont{Bosch}},
  \bibinfo{author}{\bibfnamefont{M.}~\bibnamefont{Neubert}}, \bibnamefont{and}
  \bibinfo{author}{\bibfnamefont{G.}~\bibnamefont{Paz}},
  \bibinfo{journal}{JHEP} \textbf{\bibinfo{volume}{11}}, \bibinfo{pages}{073}
  (\bibinfo{year}{2004}{\natexlab{b}}), \eprint{hep-ph/0409115}.

\bibitem[{\citenamefont{Beneke et~al.}(2005)\citenamefont{Beneke, Campanario,
  Mannel, and Pecjak}}]{Beneke:2004in}
\bibinfo{author}{\bibfnamefont{M.}~\bibnamefont{Beneke}},
  \bibinfo{author}{\bibfnamefont{F.}~\bibnamefont{Campanario}},
  \bibinfo{author}{\bibfnamefont{T.}~\bibnamefont{Mannel}}, \bibnamefont{and}
  \bibinfo{author}{\bibfnamefont{B.~D.} \bibnamefont{Pecjak}},
  \bibinfo{journal}{JHEP} \textbf{\bibinfo{volume}{06}}, \bibinfo{pages}{071}
  (\bibinfo{year}{2005}), \eprint{hep-ph/0411395}.

\bibitem[{\citenamefont{Fleming
  et~al.}(2007{\natexlab{b}})\citenamefont{Fleming, Hoang, Mantry, and
  Stewart}}]{FHMS2}
\bibinfo{author}{\bibfnamefont{S.}~\bibnamefont{Fleming}},
  \bibinfo{author}{\bibfnamefont{A.~H.} \bibnamefont{Hoang}},
  \bibinfo{author}{\bibfnamefont{S.}~\bibnamefont{Mantry}}, \bibnamefont{and}
  \bibinfo{author}{\bibfnamefont{I.~W.} \bibnamefont{Stewart}},
  \bibinfo{journal}{In preparation}  (\bibinfo{year}{2007}{\natexlab{b}}).

\bibitem[{\citenamefont{Dasgupta and Salam}(2001)}]{Dasgupta:2001sh}
\bibinfo{author}{\bibfnamefont{M.}~\bibnamefont{Dasgupta}} \bibnamefont{and}
  \bibinfo{author}{\bibfnamefont{G.~P.} \bibnamefont{Salam}},
  \bibinfo{journal}{Phys. Lett.} \textbf{\bibinfo{volume}{B512}},
  \bibinfo{pages}{323} (\bibinfo{year}{2001}), \eprint{hep-ph/0104277}.

\bibitem[{\citenamefont{Korchemsky and Marchesini}(1993)}]{Korchemsky:1993uz}
\bibinfo{author}{\bibfnamefont{G.~P.} \bibnamefont{Korchemsky}}
  \bibnamefont{and}
  \bibinfo{author}{\bibfnamefont{G.}~\bibnamefont{Marchesini}},
  \bibinfo{journal}{Phys. Lett.} \textbf{\bibinfo{volume}{B313}},
  \bibinfo{pages}{433} (\bibinfo{year}{1993}).

\bibitem[{\citenamefont{Neubert}(2004)}]{Neubert:2004dd}
\bibinfo{author}{\bibfnamefont{M.}~\bibnamefont{Neubert}}
  (\bibinfo{year}{2004}),
  \eprint{\href{http://arXiv.org/abs/hep-ph/0408179}{hep-ph/0408179}},
  \urlprefix\url{http://arXiv.org/abs/hep-ph/0408179}.

\bibitem[{\citenamefont{Korchemsky and Radyushkin}(1987)}]{Korchemsky:1987wg}
\bibinfo{author}{\bibfnamefont{G.~P.} \bibnamefont{Korchemsky}}
  \bibnamefont{and} \bibinfo{author}{\bibfnamefont{A.~V.}
  \bibnamefont{Radyushkin}}, \bibinfo{journal}{Nucl. Phys.}
  \textbf{\bibinfo{volume}{B283}}, \bibinfo{pages}{342} (\bibinfo{year}{1987}).

\bibitem[{\citenamefont{Gardi}(2000)}]{Gardi:2000lr}
\bibinfo{author}{\bibfnamefont{E.}~\bibnamefont{Gardi}},
  \bibinfo{journal}{JHEP} \textbf{\bibinfo{volume}{04}}, \bibinfo{pages}{030}
  (\bibinfo{year}{2000}), \eprint{hep-ph/0003179}.

\end{thebibliography}

\end{document}